\documentclass[onecolumn]{aa}
\usepackage{graphics}
\usepackage{graphicx}
\usepackage{amsfonts}
\include{epsf}
\include{epstopdf}
\DeclareGraphicsExtensions{.eps,.jpg,.pdf}
\begin{document}

\newcommand{\beq}{\begin{equation}}
\newcommand{\eeq}{\end{equation}}
\def\sgn{{\;{\rm sgn}}}
\def\sech{{\;{\rm sech}\;}}
\def\la{\hbox{\raise.35ex\rlap{$<$}\lower.6ex\hbox{$\sim$}\ }}
\def\ga{\hbox{\raise.35ex\rlap{$>$}\lower.6ex\hbox{$\sim$}\ }}
\def\beqa{\begin{eqnarray}}
\def\eeqa{\end{eqnarray}}
\def\sub#1{_{_{#1}}}
\def\order#1{{\cal O}\left({#1}\right)}
\newcommand{\sfrac}[2]{\small \mbox{$\frac{#1}{#2}$}}
\def\half{\sfrac{1}{2}}

\title{{The Equations of Magnetoquasigeostrophy}}
\author{O. M. Umurhan\inst{1-2}}

   \offprints{O.M. Umurhan \email{oumurhan@ucmerced.edu}}

   \institute{
   School of Physics and Astronomy, Queen Mary
   University of London, London E1 4NS, U.K.\
     \and
        School of Natural Sciences, University of California Merced,
      Merced, CA 95343, USA\
}

\date{}

\abstract
     {The dynamics contained in magnetized layers of exoplanet atmospheres are important to understand in order to characterize what observational signatures they may provide for future observations.  It is important to develop a framework to begin studying and learning the physical processes possible under those conditions and what, if any, features contained in them may be observed in future observation missions.}
     {The aims of this study is to formally derive, from scaling arguments, a manageable reduced
     set of equations for analysis, i.e. a magnetic formulation  
     of the equations of quasigeostrophy appropriate for a multi-layer atmosphere. We check these derived
     equations for consistency with respect to similar equations currently used in the literature like the magnetized shallow water equations and their precursors.  The main goal is to provide a simpler theoretical platform to explore the dynamics possible within confined magnetized layers of exoplanet atmospheres. }
     {We primarily use scaling arguments to derive the reduced equations of ``magnetoquasigeostrophy" which assumes 
     dynamics to take place in an atmospheric layer which is vertically thin compared to its horizontal scales.  Furthermore, the derivation exploits the fact that the Rossby Numbers of the emergent flows are small and that the Cowling Number is also near $1$, the latter of which measures the relative content the
     energy per unit volume contained in the magnetic field 
      to the corresponding kinetic energy of the flow.}
     { The magnetized incarnation of the quasigeostrophic equations for a two-layer system are fully rederived
from scaling arguments.  The resulting equation set retains features existing in standard shallow-water magnetohydrodynamic equations but are absent in more classical derivations of the quasi-geostrophic limit, namely, the non-divergence of the in-plane components of the magnetic field.  We liken this non-divergence of the in-plane magnetic fields as indicative of a quantity whose behaviour mimics a  
two-dimensional ``pseudo"-magnetic monopole source.  
We also find, using the same scaling argument procedures, appropriate limits of the fundamental parameters of the system which yield reduced equations describing the flow dynamics
primarily characterized by magnetostrophic balance. }
    {The standard scaling arguments employed here show how traditional magnetized quasigeostrophic 
    equations connect to their magnetized shallow water forms.  The equations derived are amenable to 
    analysis using well-known techniques.}

\titlerunning{Equations of Magnetoquasigeostrophy}

\keywords{Hydrodynamics, Magnetohydrodynamics, Exoplanets}

  \maketitle
  
\section{Introduction}
It is now generally accepted that hot extrasolar giant planets (``hot-Jupiters") are commonplace in the Galaxy.
  Many orbit their parent stars in so close that recent calculations by Koskinen et al. (2010) indicate that the upper atmospheric layers of these hot-Jupiters are sufficiently irradiated by the parent star's UV radiation field (and therefore sufficiently ionized) to treat the gas as a plasma.  It is important
therefore to understand what sorts of observational indications these upper layers may show to future telescope missions.
Model atmosphere calculations show that these upper layers are likely to be
stably stratified (like the Earth's stratosphere)
and, therefore, it is justifiable to use global circulation models to describe
and explore their flow dynamics (Cooper \& Showman 2002, Cho et al. 2003, Menou \& Rauscher 2009, to name
only a few). 
Preliminary examinations of the response of magnetized shallow-water models of exoplanets reported
by Cho (2008)
indicate that under not-too unreasonable conditions, the atmospheres of irradiated exoplanets could settle into a collection of stationary, planetary scale, magnetized vortices with zero total circulation.  
If this turns out to be a robust feature for hot irradiated Jupiters (for instance) then there will be obsevational consequences which may be detectable by the next generation of space missions.

\par
The aim of this work is to begin to formulate a systematic approach
toward understanding the dynamics of stably stratified flows on exoplanets under the influence of magnetohydrodynamic
effects which are usually absent in standard meteorological modeling.
The simplest place to begin exploring these kind of dynamics is in a quasi-geostrophic (``QG" for short) framework
which is a setting representing dynamics (i) occurring on synoptic scales (horizontal scale dynamics which
are much smaller than the planetary radius), (ii) described by order 1 Burger numbers and (iii) characterized by small Rossby number flows.  The last of these
is a measure of the ratio of the typical rotation time of the planet to the circulation time of a
synoptic scale vortical structure.  The Burger number is also a ratio between
 the planet's rotation time to the gravity wave propagation
time across the synoptic scale.
If such an atmosphere is also magnetic, then another important number governing the quality of
the dynamics will be
the Cowling number which measures the
the relative importance of
the Lorentz force.
The Cowling Number here will be understood as the ratio of the energy contained in the magnetic field to the kinetic energy in vortical motions.
\par
When the Cowling number becomes an order 1 quantity QG dynamics will be substantially modified.
Examination of
 magnetohydrodynamic modified quasigeostrophic flow goes back to the original series of
studies in the thesis work of Gilman (Gilman 1967a-c) motivated by the problem of solar
differential rotation.
The model equations derived in Gilman's original studies are the magnetohydrodynamic analog of
the classical quasigeostrophic equations
on a $\beta$-plane.
In these equations the magnetic field is strictly two-dimensional and horizontal,
where it is assumed that the magnetic energy content contained in this layer
is outweighed by the energy contained in the hydrostatic configuration.  As a result, the in-plane
magnetic
field does not figure into the lowest order geostrophic balance.  Furthermore, the field lines,
although they behave two-dimensionally, can generate vertical field lines through the action
of vertical layer undulations.  The variation scales of these vertical
 field fluctuations are, however, small compared to the horizontal divergences (at next order) and, thus,
 though vertical field lines may be generated they play no dynamical role in the equations
 of motion at lowest order.\par
 These equations also support Rossby waves and Alfven waves, often referred to by their hybridized forms
 as `hydromagnetic-planetary' waves (Acheson \& Hide 1973).  Unlike the situation in classical QG, however,
 the magneto-quasigeostrophic equation (MGQ)
 set cannot be characterized by a conserved quantity like the potential vorticity in QG.  However, the
 magnetic potential function (sometimes called a ``flux"-function) is a materially conserved quantity of these equations.
 Also, motivated by meteorological modeling, Gilman (1967b) extended the
 basic MGQ formalism onto a multi-layer ``stacked" model where the density in each layer are constant
 possibly differing from each other.
 As it is one of our goals to formulate
 something similar, Gilman's study shall serve as a guiding procedure in this respect.

  The magnetohydrodynamic version of classical shallow water equations
 (hereafter SWMHD)
 was developed in Gilman (2000)
 in order to address the problem of the solar tachocline (Spiegel \& Zahn, 1992).
 The SWMHD
 are model equations confined to a thin-shell of magnetized gas which is stably (or marginally)
 stratified.  In the familiar non-magnetized setting, the shallow water equations retain the action
 of gravity waves while the more simplified QG models do not.  Gravity waves are absent in the latter
 because the Burger numbers are order 1 in QG, and since the timescales for dynamics
 in QG are
much longer than the local planetary rotation time (because the Rossby number is small),
it means that gravity waves are filtered out.
 In this sense, one may regard the shallow
 water equations as containing ``more" physics over the QG set.  The same is true of the SWMHD model
 in that it retains the hydromagnetic planetary waves found in the MGQ set and it further
 supports magnetically modified gravity waves whose properties and structure has received
  recent scrutiny (Schecter et al. 2001, Zaqarashvili et al. 2007, Heng \& Spitkovsky 2009).\par
  Besides the absence of magnetically modified gravity waves, the SWMHD equations and the
  original MGQ models
  also diverge from one another in content
  because the horizontal divergence of the magnetic fields
 are not zero in SWMHD as they are in the original classical MGQ
 equations.
 This means that, for example, the
 dynamical evolution of the horizontal fields cannot be formulated in terms of the evolution
 of a scalar `flux-function' as is usually done for horizontal-divergence free fields.
 It is shown in this study that
 one can view this
 non-divergence of the horizontal fields as corresponding to the presence of a globally
 conserved, two-dimensionally distributed ``pseudo"-magnetic monopole charge, $q^{(m)}(x,y,t)$.  
\footnote{This unfortunate tongue-in-cheek description is used for the lack of a better expression
  to describe how the non-zeroness of the horizontal divergence of the magnetic field would ``appear" as a monopole charge source, albeit, in two-dimensions.  Of course no such thing exists as the non-vanishing feature of the horizontal divergences of the magnetic fields means only that there correspondigly exists a vertical field with non-zero vertical gradient.} 
  Therefore, in this sense, one may characterize the solutions contained within the various model approximations according to whether or not $q^{(m)}$ is zero.  For example, the MGQ developed in the original papers of Gilman from 1967-8
 have $q^{(m)} =0$ while the SWMHD models have $q^{(m)} \neq 0$ in general.\par

\medskip
 { \em What is achieved in this work:}
Since classical
 QG offers a sound
 conceptual platform to focus on and understand the dynamics of (mainly) small Rossby number vortical flows,
 it would be logical to start studying magnetized exoplanet atmospheres from a similar vantage point.
 It makes sense to begin this exploration from a MGQ framework and, once having uncovered some of the
  dynamics that this set of equations contain, verifying that some facet of these
 dynamics are contained in the SWMHD equations.
 However, some ambiguity persists as to the fate of the so-called ``pseudo"-magnetic monopole distribution in the
 reduced MGQ system as it is identically zero in Gilman's original studies but clearly present in
 SWMHD.  Does the MGQ framework contain some features of this ``pseudo"-magnetic monopole field  $q^{(m)}$?
 The answer is yes and it is one of our goals here to
 develop an intermediate framework to study magneto-vortical dynamics free
 of gravity waves but not necessarily free of the
  $q^{(m)}$ field.  In this work the MGQ equations for arbitrary number of vertical layers
  of constant density is
 re-derived using familiar scale analysis techniques as, for instance, found in Vallis (2006) (see also Pedlosky 1987).  We display the
 derived equations for a two-layer model.
 We show how the various aforementioned features between equation sets relate to one another.
 \par
 The procedure implemented to derive the MGQ equations are sufficiently general also
  to establish the scalings which lead to magnetostrophic balance (Acheson \& Hide, 1973).
 Magnetostrophy is a limiting form for small Rossby numbers in which dominant dynamical balance
 exists between Coriolis, pressure and Lorentz forces.  The magnetostrophic approximation has
 been extensively explored to study the Earth's geodynamo and associated
 turbulence (e.g. Fearn 1997, Moffat 2008).
 In our analysis and overall setting we find that the magnetostrophic balance at low Rossby numbers is
 achieved when (i) the Cowling number is sufficiently large in proportion to the inverse Rossby number
 and (ii) the Burger number is small in proportion to the Rossby number.  After establishing this
  fundamental magnetostrophic balance we proceed to derive a set of self-contained equations describing dynamics which are analogous to the MGQ set.  We only derive the resulting equations here and shall return to
  a more thorough discussion of them in a follow-up study.\par

 In Section 2 the equations and assumptions are presented.  In Section 3 the equations of motion are expanded
 around a given latitude, scaled and analyzed.  In this section we formally define the non-dimensional
 parameters of the system and scale the equations of motion. Most importantly, we highlight the fundamental relationships that must be
 met between the various parameters in order to attain the aforementioned MGQ and magnetostrophic balancing.
 The actual expansion procedure completing the derivation of the MGQ equations is detailed in Section 4
 while the same is done for the magnetostrophic set in Section 5.
 Because the main focus of this study is about the MGQ equations, in Section 6 they are summarized
 followed by a brief discussion demonstrating
    how a MGQ model reduction does exist which retains the $q^{(m)}$ field and how, most importantly, the classical MGQ system is recovered when $q^{(m)}$ is set to zero.\par

\section{Equations and assumptions: an overview}
The general equations of motion of an ideal, incompressible MHD fluid in a frame rotating
 with rotation vector ${\bf \Omega}$ are
\beqa
\frac{d {\bf U}}{dT} + 2{\mathbf \Omega} {\mathbf{{\rm curl}}} {\bf U} &=& \frac{1}{\rho}\mathbf{{\rm{grad}}} P - g{\bf {\hat r}} + \frac{1}{\rho}{\bf J}\mathbf{{\rm curl}} {\bf B}, \\
\mathbf{{\rm div}}\cdot {\bf U} &=& 0, \label{incompressibility_proper} \\
\frac{\partial {\bf B}}{\partial T} + \bf{{\rm curl}}({\bf U} \ \mathbf{{\rm curl}} {\bf B}) &=& 0, \\
\mathbf{{\rm div}}\cdot {\bf B} &=& 0,
\eeqa
where ``div", ``grad" and ``curl" are the corresponding three dimensional operations
of divergence, gradient and curl respectively. These equations will be considered in the next section
in a Cartesionized representation around a point at some latitude located suitably away from the planet's equator.
In the Cartesionized frame, $X$ represents the zonal (longitudinal/east-west) direction while $Y$ represents the meridional (latitudinal/north-south) direction while $Z$ represents
the vertical (i.e. planetary radius) direction.
 The corresponding velocities are (respectively)
$U,V,W$ while the magnetic fields are
$B_X,B_Y,B_Z$.  $P$ represents
the pressure.
We consider the dynamics of these equations subject to the following setting
and restrictions:\par
\begin{enumerate}
\item The planetary $\beta$-plane, nominally centreed on a latitudinal zone which is significantly away from the planet's equatorial zone.
\item The dynamics of small Rossby number disturbances (i.e. Ro $\ll 1$) are of interest so that we may develop the equations as a power series in this quantity  .
\item As is standard practice in atmosphere modeling, we break the atmosphere up into several layers.
We assume that the background density fields in each vertical layer is constant so that each layer is effectively incompressible as implied by (\ref{incompressibility_proper}). The layers, however, may have differing densities
    each given by $\rho_i$ where $i$ denotes the layer under consideration.
\item We shall allow for either none, some  or all of the layers to be electrically conducting so that MHD will be a good description of the dynamics in those
    layers that are magnetically active.  This is why the induction equations for ideal MHD are included.
\end{enumerate}
\par
In the majority of the analysis performed in this study the dynamics 
are treated in two-layers. However, for the sake of generality of the procedures
we implement, let us consider the possibility of an arbitrary number of constant-density layers with,
as yet, unassigned conductivity. 
Let us suppose that the
heights delineating the transition from one constant
 density layer to another is given by ${\cal H}_1 < {\cal H}_2$
respectively where $Z=0$ represents the nominal ``bottom" of the atmosphere (see Figure 1).
Then the prescription for the current ${\bf J}$ will be represented in this way by
\beq
{\bf J} =  \varphi {\mathbf{{\rm curl}}} {\bf B}; \qquad \varphi = \left\{
\begin{array}{lr}
\varphi_N, & \ \ {\cal H}_{N-1} < Z \le {\cal H}_{N} \\
\vdots & \vdots \\
\varphi_i, & \ \ {\cal H}_{i-1} < Z \le {\cal H}_{i} \\
\vdots & \vdots \\
\varphi_1, & \ \ 0 \le Z \le {\cal H}_1
\end{array}\right.
\eeq
The functions $\varphi_i$ will take on either the values $0$ or $1$ depending upon
the layer $i$.\par
 Sections 3-4 detail the derivation of the MGQ equations with
explicit presentation of a two-layer model.
The derivation follows the standard procedure outlined in Vallis (2006, $\S$ 5.3, pgs. 207-215) while exploiting some of the scaling arguments invoked by Gilman (1967a).
 We introduce here some of the relevant scalings and corresponding non-dimensional
 parameters appearing in the
 system analyzed.  The planetary rotation as viewed from the latitude in question is scaled by $\Omega_0$\footnote{If the planetary rotation is given by $\bar\Omega$ and
if one is at latitude $\lambda$ then the rotation normal to that latitude is $\Omega_0 =\bar\Omega \sin \lambda$}.  If the typical meridional/zonal velocity scales are given by ${\cal U}$ then we may define the Rossby number
\[ {\rm Ro} \equiv \frac{\cal U}{f_0 {\cal L}},
 \]
 where $f_0 =2\Omega_0$ and $R$ is the planetary radius and ${\cal L}$ represents the
 horizontal ``synoptic" scale of the dynamics.
 Dynamic flow timescales are assumed to be 1/Ro times
longer than the Coriolis timescale $1/f_0$ - this is the basis of the quasigeostrophic analysis.
The magnetic field strength is scaled by ${\cal B}$.  This
then leads to a natural non-dimensional
quantity
\[
C \equiv \frac{{\cal B}^2}{4\pi\tilde\rho}\frac{1}{{\cal U}^2},
\]
also known as the Cowling number.
The scale density $\tilde\rho$ is equated to the density of the lower atmosphere layer.
$C$ is related to the inverse square of the usual $\beta$ parameter frequently
referred to in plasma physics studies.  The Cowling number can be understood as a measure
of the energy contained in the magnetic field versus that contained in kinetic motions.
It can also be understood as the square of the magnetic Mach number,
i.e. $C =  {{\cal U}_A}^2/{\cal U}^2$,
since ${\cal U}_A^2 \equiv {{\cal B}^2}/{4\pi\tilde\rho} $ is the Alfv\'{e}n speed squared.
\par
The atmosphere also has the following vertical length scales that are important:
 the overall vertical extent
of the atmosphere ${\cal H}$, and the vertical length scale
of atmospheric {\textit fluctuations} $\tilde h$.
Therefore, the zonal and meridional dynamical lengths ($X,Y$) are scaled by $\cal L$ while the
vertical scales ($Z$) are characterized by ${\cal H}$. Given that the flow
speeds are $\order{\cal U}$
it follows that their typical characteristic timescales ($T$) are Ro$^{-1}$ times
longer than $1/f_0$, consistent with the assertion made above.  These are all formally reintroduced
in the discussion appearing in the next section.
Finally, the Burger number, Bu, measuring the relative importance of gravity,
is given by the relationship
\[
{\rm Bu}^2 = \frac{g{\cal H}}{4\Omega_0^2 {\cal L}^2}.
\]

\section{Scalings and analysis}\label{MGQ_derivation}
The local Cartesionization of the equations of motion around a latitude $\lambda_0$
is given by,
\beqa
\frac{d {\bf U}}{dT} + 2{\Omega} \hat {\bf z}{\times} {\bf U} &=& \frac{1}{\rho}\nabla \Pi - g{\bf {\hat z}} + \frac{1}{4\pi\rho} ({\bf B}\cdot \nabla){\bf B}, \\
 \nabla\cdot{\bf U} &=& 0, \label{incompressibility} \\
\frac{d {\bf B}}{d T}   &=& ({\bf B}\cdot \nabla){\bf U}, \\
\nabla\cdot {\bf B} &=& 0,
\eeqa
where the total pressure $\Pi$ is given by
\beq
\Pi = P + \frac{1}{8\pi}{\bf B}^2.
\eeq
The coordinate $Z$ coincides with the effective gravity direction.  The component of gravity $g$ will
be taken to be constant.  In this construction  $X,Y$ denote the zonal and meridional coordinates respectively.

Before analyzing these equations on the Cartesion plane, we
explicitly nondimensionalize all quantities appearing in the equations of
motion according to the dimensional quantities stated above.
Henceforth, all lower case Latin symbols (some with hats over them) denote the corresponding non-dimensional
quantity.  The primary parameter that will be used for the following expansions
will be the Rossby number, Ro.
In terrestrial and planetary
studies the Rossby number is typically substantially less than one, i.e.
Ro $\ll 1$.
${\cal U}$ is the typical velocity scale of the horizontal planetary scale atmospheric motions, that is to say,
\beq
U \rightarrow {\cal U} u, \qquad V \rightarrow {\cal U} v,
\eeq
where $u,v$ are the non-dimensionalized order one representations of the
longitudinal and latitudinal velocities.
Since ${\cal U}$ measures an effective overturning speed of the synoptic scale
vortex structures in the atmosphere, the corresponding time scale associated
with it scales as Ro$/\Omega_0$.
Thus we scale the horizontal lengths and
time according to:
\[
X \rightarrow {\cal L} x; \quad Y \rightarrow {\cal L} y, \quad T \rightarrow t/({\rm Ro}\Omega_0)
\]
where $x,y,t$ are now understood to be non-dimensionalized quantities.
The vertical scale is measured by ${\cal H}$ and this will be considered
small compared to ${\cal L}$.  In problems in which the atmosphere is broken
up into subplayers ${\cal H}$ represents the full vertical extent of the whole
atmosphere under consideration.
In order to recover the quasi-geostrophic ordering
the smallness of the ratio ${\cal H}/{\cal L}$
needs only be $\ll 1$.  Correspondingly, the vertical velocity will
also be assumed to be small in proportion to the ratio $\cal{H/L}$.  Thus we
scale these two quantities accordingly as
\[
Z \rightarrow {\cal H} z, \qquad W \rightarrow \left(\frac{{\cal H}}{{\cal L}}\right) {\cal U} w
\]
where $z$ and $w$ are order 1 non-dimensional quantities.  The variation of the
Coriolis parameter (which is the usual function of latitude on a planet)
is written as
\[
\Omega = \Omega_0\left(1 + \frac{{\cal L}}{R} \beta y + \cdots \right),
\]
where $\Omega_0$ is the evaluation of the projected planetary rotation
at latitude $\lambda_0$ in which $y {\cal L}/R \equiv \lambda - \lambda_0$.
$\Omega_0 = \bar\Omega \sin \lambda_0$ and
$\beta = \cos\lambda_0$, where $\bar\Omega$ is the planetary rotation
rate.  The ratio ${\cal L}/R$ will be formally shown to be $\order{\rm Ro}$ in Section 4 in
order to retain the planetary-$\beta$ effect.
The magnetic field strength will be measured by ${\cal B}$.  Thus, in keeping
with the formalism developed by Gilman (2000), the horizontal fields are
scaled by this,
\[
B_x \rightarrow {\cal B} b_x, \qquad B_y \rightarrow {\cal B} b_y,
\]
while, because of the small aspect ratios under consideration, \emph{we shall
only consider} vertical
fields which are smaller than the horizontal components in proportion
to the ratio $\cal{H/L}$,
\[
B_z \rightarrow \frac{{\cal H}}{{\cal L}} b_z.
\]
The vertical extent of the atmosphere is subdivided into sublevels with vertical
scale ${\cal H}_i$ where $i$ labels the level number.  For our considerations
we assume that all $\Delta {\cal H}_i = {\cal H}_{i} - {\cal H}_{i-1}$, representing the thicknesses
 of each layer, are an order 1 fraction of the overall vertical
scale ${\cal H}$.  We let $\tilde h_i$ represent perturbations about the
mean level height ${\cal H}_{i0}$, i.e.,
\[
{\cal H}_i = {\cal H}_{i0} + \tilde h_i,
\]
If we assume that the deviations scale by a characteristic length $\tilde h$, then
we may introduce another parameter  $\delta$
which measures the the deviations against the vertical scale of the atmosphere
\beq
\delta \equiv \frac{\tilde h}{\cal H}.
\eeq
In typical quasigeostrophic scalings this factor $\delta$ is assumed small, usually
$\order{{\rm Ro}}$ for Ro small.
\subsection{Analysis of the vertical momentum equation}

We seek to develop reductions of the equations of motion which are dynamically
in hydrostatic equilibrium.
This means
assuming the pressure scaling to be
\[
\Pi \rightarrow g{\cal H}\tilde\rho\breve\Pi,
\]
where $\tilde\rho$ is the density scale, typically of the lowest atmosphere layer.
Examining the vertical component of the momentum equation
with the scalings thus far presented reveals
\beq
{\rm Ro}^2 \left(\frac{{\cal H}}{{\cal L}}\right)^2\left(\partial_t + {\bf u}\cdot\nabla
+w\partial_z
\right){ w} =
{\rm Bu}^2
\frac{\tilde\rho}{\rho}\left(\partial_z \breve \Pi +\frac{\rho}{\tilde\rho}\right)
+{\rm Ro}^2 C \frac{\tilde\rho}{\rho}\left(\frac{{\cal H}}{{\cal L}}\right)^2({\bf b}\cdot\nabla + b_z\partial_z){ b_z}
.\label{vert_mom}
\eeq
Where the Burger number, Bu, and the Cowling number, $C$, are as defined at the end of
Section 2.
To insure that hydrostatic balance is always dominant we shall assume
 Bu always dominates the other terms appearing in (\ref{vert_mom}).  This means
 to say specifically that unless
 \beq
 {\rm Bu}^2 \gg {\rm Ro}^2 \left(\frac{{\cal H}}{{\cal L}}\right)^2 , \quad
 {\rm and} \quad
 {\rm Bu}^2 \gg {\rm Ro}^2 \left(\frac{{\cal H}}{{\cal L}}\right)^2 C,
 \eeq
 the assumption of hydrostatic balance will fail.
 For planetary atmospheres this criterion is generally met
 since ${\cal H}/{{\cal L}} \ll 1$.  The new piece here is the second disparity
 involving the Cowling number.  We return to this shortly below.
As of this point we have said nothing about the specific orderings
required of the both the Burger and Cowling numbers.  In general for this
study we shall
assume that Ro and the ratio ${\cal H}/{\cal L}$ are sufficiently smaller than 1.
 It follows that the lowest order balance of the vertical momentum
equation (\ref{vert_mom}) is hydrostatic,
\beq
\partial_z \breve \Pi +\frac{\rho}{\tilde\rho} = 0.
\eeq
Henceforth we shall treat $\rho$ as being non-dimensionalized by $\tilde \rho$ which
means, in practice, that $\rho/\tilde\rho$ will be replaced everywhere by $\rho$.
When multiple layers are handled we shall characterize the density as being constant
in each layer with value $\rho_i$.  Furthermore,
to be consistent with what was stated earlier, the density of the lowest layer will always
be unity, thus $\rho_1 = 1$.
The procedure performed here
essentially mimics the strategy employed in Vallis (2006).  We allow the
  top layer to move about freely and assume the pressure is zero on the top surface.
  The solution for the pressure for an N-layered atmosphere (with a flat
   bottom located at $z=0$) expressed in terms of the non-dimensionalized
  lid coordinates $H_i$ is given by
\beq
\breve \Pi = \int_{H_N}^z \frac{\rho}{\tilde\rho} dz = \left \{
\begin{array}{cc}
 \rho_{N}(H_{N} - z),
& {H_N} > z \ge H_{N-1} \cr
\sum_{k=N}^N \rho_k(H_k-H_{k-1}) +
 \rho_{N-1}(H_{N-1} - z), \ \
& {H_{N-1}} > z \ge H_{N-2} \cr
\vdots & \vdots \cr
\sum_{k=i+1}^N \rho_k(H_k-H_{k-1}) +
 \rho_{i}(H_{i} - z),
& {H_i} > z \ge H_{i-1} \cr
\vdots & \vdots \cr
\sum_{k=2}^N \rho_k(H_k-H_{k-1}) + \rho_1(H_1-z), & H_1 \ge z \ge 0.
\end{array}
\right .
\eeq
Note that in the above formulation the summation procedure begins with layer
$i = N-1$ and that it is explicitly absent in the top, $i=N$, layer.
As was motivated earlier, the levels $H_i$ are further written in the form
\beq
H_i = H_{i0} + \delta h_i,
\eeq
where the set $\{H_{i0}\}$ are constants.
Henceforth
the procedure
outlined in the remainder of this study will focus only on a two-layer system with
the understanding that it straightforwardly generalizes to a multi-layer configuration.
\footnote{To avoid any ambiguity: the general form of the disturbance pressure would
be given by the general formulae
$\hat\Pi_i^{(0)} = \rho_{i}h_{i} +
\sum_{k=i+1}^N \rho_k(h_k-h_{k-1})$ with $\hat\Pi_N^{(0)} = \rho_n h_n$.}
 Given the functional form postulated above
we find that horizontal gradients ($\nabla \equiv {\hat {\bf x}}\partial_x +
{\hat {\bf y}}\partial_y$)
of the pressure are given by
\beq
\nabla\breve \Pi = \delta\nabla\hat \Pi = \delta \nabla
\Biggl \{
\begin{array}{lr}
\hat \Pi_2 = \rho_2 { h_2}, & {H_2} > z \ge H_1 \cr
\hat \Pi_1 =\rho_2 h_2 +  (\rho_1-\rho_2) h_1  , & H_1 \ge z \ge 0.
\end{array}
\label{hat_pi_definition}
\eeq
\subsection{Analysis of the horizontal momentum equations}
Now we are prepared to move onto the horizontal momentum equations.  With the scalings
proposed above the equations become
\beqa
&  & {\rm Ro}^2\left(\partial_t + {\bf u}\cdot\nabla + w\partial_z\right){\bf u}
+ 2{\rm Ro}\left(1+\frac{{\cal L}}{R}\beta y\right)\hat {\bf z}\times{\bf u} = \nonumber \\
& & \hskip 2.0cm
-\delta {\rm Bu}^2
\frac{1}{\rho}\nabla\hat\Pi
+{\rm Ro}^2\frac{C}{\rho}({\bf b}\cdot\nabla + b_z\partial_z){\bf b} + \order{{\rm Ro}^2 \frac{{\cal L}}{R}},
\label{horizontal_equation}
\eeqa
in which ${\bf u} = u {\hat{\bf x}} + v {\hat{\bf y}}$
and  ${\bf b} = b_x {\hat{\bf x}} + b_y {\hat{\bf y}}$.  We note that
the curvature terms neglected at this stage come in at an ordering
of $\order{{\rm Ro}^2} \cdot \order{{\cal L}/R}$, and since ${\cal L}/R$ will be
assumed to be $\ll 1$, the curvature terms will be neglected throughout.
It is at this stage that we explore various balances:

\subsubsection{Quasigeostrophic/magnetoquasigeostrophic scalings}\label{MGQ_scalings}
We assume that the deviations from the mean height
scale to the Rossby number, i.e.
$\delta  = \order{{\rm Ro}}.$
Although it would be enough to require that ${\cal L}/R \ll 1$, in order
to include the planetary $\beta$ we require that
$
{\cal L}/{R} \sim\order{{\rm Ro}}.
$
In Section \ref{detailed_MGQ} we detail the full procedure leading to the
quasigeostrophic reduction, however, we show here the leading order balances
to illustrate the main flavour of the results.
Expanding the pressure and velocities within each layer $i$ in powers of Ro according to,
$\hat\Pi_i = \hat\Pi_i^{(0)} + {\rm Ro} \hat\Pi_i^{(1)} + \cdots$,
$ \ {\bf u}_i = {\bf u}_i^{(0)} + {\rm Ro} {\bf u}_i^{(1)} + \cdots$ and ${\bf b}_i = {\bf b}_i^{(0)} + {\rm Ro} {\bf b}_i^{(1)} + \cdots$,
then
it follows that the leading order terms of (\ref{horizontal_equation}), which are $\order{{\rm Ro}}$,
reduce to the fundamental statement of geostrophic balance, that is to say,
\beq
2\hat {\bf z}\times{\bf u}_i^{(0)} =
-{\rm Bu}^2
\frac{1}{\rho_i}\nabla\hat\Pi^{(0)}_i.
\label{lowest_order_Geostrophy}
\eeq
Because we consider the density in each layer to be constant, taking the curl of the
above equation immediately reveals that the horizontal divergence of the lowest
order velocity field is identically zero
\[
\partial_x u_i^{(0)} + \partial_x v_i^{(0)} =0.
\]
Therefore it follows from this that the lowest order vertical velocity $w_i^{(0)}$
must zero.  Requiring $\delta$ to be less than order 1 actually follows
from the horizontal divergence free condition of the lowest order horizontal velocity field.
This consistency is formally shown in Section \ref{detailed_MGQ}.
\par
In order to understand how magnetic effects enter the resulting equations we
carry out the expansion to order Ro$^2$ to find,
\[
(\partial_t + {\bf u}_i^{(0)}\cdot\nabla ){\bf u}_i^{(0)}
+ 2\hat {\bf z}\times{\bf u}_i^{(1)}
+ \beta y\hat {\bf z}\times{\bf u}_i^{(0)}
 =
-{\rm Bu}^2
\frac{1}{\rho_i}\nabla\hat\Pi_i^{(1)}
+\frac{C}{\rho_i}({\bf b}_i^{(0)}\cdot\nabla){\bf b}_i^{(0)}.
\]
The above is a restatement of Eq. (\ref{horizontal_equation_next_order}) where in
arriving at this equation
it is argued that the horizontal components of the magnetic field are independent of $z$
at lowest order.  All but the
last term on the RHS of the above expression constitute the next order corrections
which ultimately lead to the classical equations of quasigeostrophy.  In particular
if ${\cal C} \ll 1$ then this limiting form identically leads to classical QG.
We see that the first non-trivial inclusion of magnetic effects comes in when
${\cal C}$ is an order 1 quantity.  In this event, the resulting equations will
be called {\emph{magnetoquasigeostrophy}} (MGQ).  The completed derivation of the latter
is found in Section \ref{detailed_MGQ}.

\subsubsection{Magnetostrophic balance}
When magnetic effects are relatively strong there exists another more general lowest
order balance between Coriolis, horizontal pressure gradients and the
horizontal components of the Lorentz force.  This {\em magnetostrophic balance}
has been explored in the context of the Earth's geodynamo
(Acheson \& Hide 1973, Fearn 1997, Moffatt 2008, to name just a few).
Inspection of (\ref{horizontal_equation})
shows that in order to bring the Lorentz term in on the same order as the
Coriolis term requires $C = \order{{\rm Ro}^{-1}}$.  Thus, we
define a new number
\[
{A} \equiv \frac{{\cal B}^2}{8\pi\tilde\rho {\cal U} \Omega_0 {\cal L} },
\]
which we shall, henceforth, call the {\emph{Acheson Number}}.  For this three-way balance
to occur the Acheson Number must be an order 1 quantity.  Similar inspection of
(\ref{horizontal_equation}) shows that in order for the horizontal pressure
gradient to enter the mix we must have the combination $\delta \cdot$Bu$^2=
\order {\rm Ro}$.  Inspection also shows that if the resulting scaling arguments are consistent
with the resulting equations $\delta$ must be $\order 1$ (see below).  Thus we define a modified Burger number
$\tilde {\rm Bu}$ such that Bu = Ro$^{1/2}\tilde {\rm Bu}$, where
\[
\left(\tilde {\rm Bu}\right)^2 \equiv \frac{gH}{2\Omega_0 {\cal L} {\cal U}}.
\]
With the assumption that (i) all quantities may be expanded in powers of Ro (as we did in the previous section) and
(ii) the lowest order horizontal velocity and magnetic
field components are $z$-independent
in a layer,
then we find that the magnetostrophic balances in the horizontal directions are
\beq
2\hat {\bf z}\times{\bf u}_i^{(0)} =
-\tilde{{\rm Bu}}^2
\frac{1}{\rho_i}\nabla\hat\Pi_i^{(0)} + \frac{{\rm A}}{\rho_i}
({\bf b}_i^{(0)}\cdot\nabla){\bf b}_i^{(0)}.
\label{lowest_order_magnetostrophic_balance}
\eeq
The  horizontal divergence of the lowest order velocity is not zero in general. In fact,
\beq
\partial_x u_i^{(0)} + \partial_x v_i^{(0)} = \frac{{\rm A}}{2\rho_i}\left(\partial_x b_{xi}^{(0)} +
\partial_x b_{yi}^{(0)} +
{\bf b}_i^{(0)}\cdot\nabla\right)J_i^{(0)} \neq 0,
\label{horizontal_divergence_MS}
\eeq
in which $J_i^{(0)}$ is the lowest order vertical current in the layer, $J_i^{(0)} \equiv
\partial_x b_{yi}^{(0)} -\partial_y b_{xi}^{(0)}$.  Unlike what we encounter
in the MGQ scalings, it follows that
the lowest order vertical velocity cannot be zero at lowest order.
Since $\delta$ measures the vertical variations of each layer height and since
the latter is related
to the vertical velocity, in order for the magnetostrophic scalings to be consistent
and to have the lowest order horizontal divergences to be non-zero
\emph{it must be} that $\delta = \order 1$, which is to say
vertical motions are driven by vertical currents due to the Lorentz force at lowest order.
The continued analysis of the magnetostrophic equations and development of
a self-contained equation set is presented in Section \ref{MSeqns}.
\par 
In Table \ref{BalancesTable}
we summarize the relative scalings of the relevant quantities that produce the
reduced equations discussed to now (i.e., QG, MQG and magnetostrophy).

\begin{table}
\caption{Summary of scalings and lowest order balances for hydrostatic disturbances. All balances 
shown assume Ro $\ll 1$. The quoted extreme values in describing magnetostrophic balance
 is the same as saying the Acheson and modified Burger numbers (denoted in the text by
$A$ and $\tilde{\rm {Bu}}$ respectively) are $\order 1$ .}
 \label{BalancesTable}
\centering
\begin{tabular}{cccccc}
\hline
{Balance Name} & $\order{{\rm Bu}}$ & $\order{\delta}$ & $\order{\tilde C}$ & $\order{\frac{{\cal L}}{R}} $
 &{Lowest Order Balance}\\
 \hline
QG & $1$ & Ro & $\ll$ 1 & Ro &  Coriolis, Pressure Gradient \\
MQG & $1$ & Ro &  1 & Ro &Coriolis, Pressure Gradient\\
Magnetostrophy   & Ro$^{1/2}$& 1 & Ro$^{-1}$ & $\ll 1$ & Coriolis, Pressure Gradient,Lorentz Force\\
\hline
\end{tabular}
\par
\end{table}

\subsection{Analysis of the incompressibility and induction equations}
The scalings assumed at the beginning of this section result in the incompressiblity equation
in each layer
rewritten as
\beq
\partial_x u_i + \partial_y v_i + \partial_z w_i = 0.
\label{in_layer_incompressibility}
\eeq
The induction equations are also similarly written as
\beq
(\partial_t + {\bf u}_i\cdot\nabla + w_i\partial_z){\bf b}_i =
({\bf b}_i\cdot\nabla + b_{zi} \partial_z) {\bf u}_i
\eeq
with the vertical field component $b_{zi}$ relating to the other field components through the
divergence free condition
\beq
\partial_x b_{xi} + \partial_y b_{yi} + \partial_z b_{zi} = 0.
\label{in_layer_sourcefree}
\eeq

\section{Detailed derivation of the equations of magnetoquasigeostrophy}\label{detailed_MGQ}
We formally develop the scaling analysis and equation reduction that we
began in Section \ref{MGQ_scalings}.
Quantities in each layer $i$ are expanded in the following way,
\beqa
{\bf u}_i &=& {\bf u}_i^{(0)} + {\rm Ro} \ {\bf u}_i^{(1)} + \cdots \nonumber \\
{\bf b}_i &=& {\bf b}_i^{(0)} + {\rm Ro} \ {\bf b}_i^{(1)} + \cdots \nonumber \\
\hat\Pi_i &=& {\Pi}_i^{(0)} + {\rm Ro} \ {\Pi}_i^{(1)} + \cdots ,
\eeqa
where the superscript denotes which order of the Ro expansion the term represents.
As we stated before, insertion of these expansions into (\ref{horizontal_equation})
results in
(\ref{lowest_order_Geostrophy}) to lowest order.
Since
we are considering a two-layer problem we have by layer that
\beqa
u_1^{(0)} &=& -\partial_y\frac{{\rm Bu}^2}{2\rho_1}\hat\Pi_1^{(0)}, \qquad
v_1^{(0)} = \partial_x\frac{{\rm Bu}^2}{2\rho_1}\hat\Pi_1^{(0)},
\eeqa
and
\beqa
u_2^{(0)} &=& -\partial_y\frac{{\rm Bu}^2}{2\rho_2}\hat\Pi_2^{(0)} \qquad
v_2^{(0)} = \partial_x\frac{{\rm Bu}^2}{2\rho_2}\hat\Pi_2^{(0)},
\eeqa
It will prove to be more convenient to write the pressure fields $\hat\Pi_i$ in
terms of a streamfunction $\psi_i$ such that
\beq
\psi_1^{(0)} \equiv \frac{{\rm Bu}^2}{2\rho_1}\hat\Pi_1^{(0)}
= \frac{{\rm Bu}^2}{2}\left[
\frac{\rho_2}{\rho_1} {} h_2^{(0)} +  \left(1-\frac{\rho_2}{\rho_1}\right) h_1^{(0)}
\right],
\eeq
and
\beq
\psi_2^{(0)} \equiv \frac{{\rm Bu}^2}{2\rho_2}\hat\Pi_2^{(0)}
=   \frac{{\rm Bu}^2}{2}{h_2^{(0)}},
\eeq
where the height deviations have been similarly expanded in powers of Ro, i.e.
\[
h_i = h_i^{(0)} + {\rm Ro} h_i^{(0)}  + \cdots
\]
This rewriting expresses the relationships more transparently
\beq
u^{(0)}_1 = -\partial_y\psi_1^{(0)},  \quad v^{(0)}_1 = \partial_x\psi_1^{(0)}, \qquad
u^{(0)}_2 = -\partial_y\psi_2^{(0)},  \quad v^{(0)}_2 = \partial_x\psi_2^{(0)}.
\eeq
Evidently the fields ${\bf u}_i^{(0)}$  are 2-dimensional incompressible flow
independent of the vertical coordinate $z$.
This means
from the incompressibility equation (\ref{in_layer_incompressibility})
\beqa
\partial_x u_i^{(0)} + \partial_y v_i^{(0)} + \partial_z w_i^{(0)} &=& 0, \nonumber
\eeqa
we get
\beqa
\partial_z w_i^{(0)} &=& 0.
\eeqa
Thus the leading order vertical velocity term is zero \footnote{if the totality
of all the layers are not moving.} motivating its
expansion to be
\beq
w_i = {\rm Ro} \ w_i^{(1)} + \cdots
\eeq
Note that this is necessarily consistent with the assumption made at the outset
that $\delta = \order{{\rm Ro}}$.  In other words this says that
if the vertical velocities are small then so must
be the vertical variations of height.\par
We turn to the leading order equation for the horizontal compoenents of the
magnetic field and we find
\[
\left(\partial_t + {\bf u}_i^{(0)}\cdot \nabla\right) {\bf b}_i^{(0)}
=
\Bigl({\bf b}_i^{(0)}\cdot\nabla + b_{zi}^{(0)}\partial_z\Bigr){\bf u}_i^{(0)}.
\]
However, since the horizontal velocity components are independent of $z$ the
above equation reduces to
\beq
\left(\partial_t + {\bf u}_i^{(0)}\cdot \nabla\right) {\bf b}_i^{(0)}
=
{\bf b}_i^{(0)}\cdot\nabla {\bf u}_i^{(0)}, \label{leading_order_induction}
\eeq
which implies that ${\bf b}_i^{(0)}$ is independent of $z$ as well.  The leading
order expansion of the source
free condition (\ref{in_layer_sourcefree})
\beq
\partial_x b_{xi}^{(0)} + \partial_y b_{yi}^{(0)} = -\partial_z b_{zi}^{(0)}
\label{bz0_relation}
\eeq
relates the leading order vertical field $b_{zi}^{(0)}$ to ${\bf b}_i^{(0)}$.  We note
that it means $b_{zi}^{(0)}$ is at most a linear function of the coordinate $z$ within
each layer.
Furthermore, and quite unlike the geostrophic balance situation we encountered
before for the leading order horizontal velocities, the horizontal magnetic fields
are not strictly two-dimensional, i.e.
$b_z^{(0)}$ is not identically zero (Gilman, 2000).  It follows from this
that there are modes of dynamical activity which will allow the divergence of the leading order horizontal magnetic field resulting in a certain amount of ``breathing" into the vertical field component.

At order Ro of the incompressibility condition we find
\beq
\partial_x u_i^{(1)} + \partial_y v_i^{(1)} = -\partial_z w_i^{(1)}.
\eeq
The remaining order Ro terms of the horizontal momentum balance equation (\ref{horizontal_equation})
are
\beqa
&  & (\partial_t + {\bf u}_i^{(0)}\cdot\nabla ){\bf u}_i^{(0)}
+ 2\hat {\bf z}\times{\bf u}_i^{(1)}
+ \beta y\hat {\bf z}\times{\bf u}_i^{(0)}
 =
-{\rm Bu}^2
\frac{1}{\rho_i}\nabla\hat\Pi_i^{(1)}
+C({\bf b}_i^{(0)}\cdot\nabla){\bf b}_i^{(0)} ,
\label{horizontal_equation_next_order}
\eeqa
where we have explicitly used the fact that ${\bf b}_i^{(0)}$ is independent
of $z$ in writing (\ref{horizontal_equation_next_order}).
Because an examination of the next order expansion of the vertical momentum balance
equation reveals that $\breve\Pi_i^{(1)}$ is independent of the vertical coordinate,
it follows that
the next order horizontal velocity corrections ${\bf u}_i^{(1)}$ are also independent
of $z$.
We may therefore take the curl of the above equation by operating the y-momentum
component by $\partial_x$ and subtracting from it the $\partial_y$ operation upon the
x-momentum equation to get
\beqa
& & (\partial_t + {\bf u}_i^{(0)}\cdot\nabla )
\left(\partial_x v_i^{(0)} - \partial_y u_i^{(0)}\right)
+ 2\Omega_i^{(1)} = \nonumber \\
& & \hskip 1.75cm C_i({\bf b}_i^{(0)}\cdot\nabla)(\partial_x b_{yi}^{(0)} - \partial_y b_{xi}^{(0)})
+ C_i(\partial_x b_{xi}^{(0)} + \partial_y b_{yi}^{(0)})(\partial_x b_{yi}^{(0)} - \partial_y b_{xi}^{(0)}).
\label{next_order_horiz_momentum}
\eeqa
with $C_i \equiv C/\rho_i$ and
where, for notational convenience, we have defined
 \beq
 \Omega_i^{(1)} = -(\partial_x u_i^{(1)} + \partial_y v_i^{(1)})
 \eeq

 In order to proceed we must determine how $\Omega_i^{(1)}$
  relate to $h_i^{(0)}$.  To do this we begin by determining the equation for each of the level
 heights $h_i$.  The previous scaling analysis leads to the non-dimensionalized version for them which, at leading order,
 reveals that
 \beq
 (\partial_t + {\bf u}_i^{(0)}\cdot\nabla)h_j^{(0)} = w^{(1)}_i(z = H_j)
 \eeq
 where care must be taken since one may evaluate the motion of the $j^{{\rm th}}$
 surface using the velocity fields from layer $i$.
 Specifically we explicitly
 write out the solution to the vertical velocities for each layer.  In the bottom
 layer we have
 \beq
 w_1^{(1)} = \Omega^{(1)}_1 z,
 \eeq
 in which we have implicitly forced the vertical velocity to be zero at the bottom $z=0$.
 To leading order it follows that
 \beq
 w^{(1)}_1(z = H_1) = H_{10} \Omega^{(1)}_1.
\eeq
When viewed from the lower layer then we have the motion of the lower layer's interface to be
given by
\beq
\left(\partial_t + {\bf u}_1^{(0)}\cdot\nabla\right) h_1^{(0)} =
H_{10} \Omega^{(1)}_1. \label{h1_Omega1}
\eeq
For the upper layer we have
\beq
w_2^{(1)} = w_{20}^{(1)} + \Omega^{(1)}_2 z,
\eeq
where $w_{20}^{(1)}$ is a constant velocity to be determined below.
Now it is evident that the motion of the bottom boundary when viewed
from the upper layer is
\beq
\left(\partial_t + {\bf u}_2^{(0)}\cdot\nabla\right) h_1^{(0)} =
w_{20}^{(1)} + H_{10}\Omega^{(1)}_2,
\eeq
while the motion of the upper boundary when viewed from the upper layer
is
\beq
\left(\partial_t + {\bf u}_2^{(0)}\cdot\nabla\right) h_2^{(0)} =
w_{20}^{(1)} + H_{20}\Omega^{(1)}_2.
\eeq
Subtracting these two expresses the evolution of the width of the
upper layer in terms of $\Omega^{(1)}_2$, that is to say,
\beq
\left(\partial_t + {\bf u}_2^{(0)}\cdot\nabla\right) \left(h_2^{(0)} - h_1^{(0)}\right) =
(H_{20}-H_{10})\Omega^{(1)}_2, \label{h2_Omega2}
\eeq
which, we note, is purely in terms of the velocity fields ${\bf u}_2^{(0)}$.
We are now in a position to rewrite
(\ref{next_order_horiz_momentum}) layer by layer by replacing all
instances of $\Omega^{(1)}_i$ with its appropriate
expressions in terms of $h^{(0)}_i$ given by (\ref{h1_Omega1}) and
(\ref{h2_Omega2}).  To this end, we begin with
the bottom layer
\beqa
& & \left(\partial_t + {\bf u}_1^{(0)}\cdot\nabla \right)
\left[\partial_x v_1^{(0)} - \partial_y u_1^{(0)} - \frac{2h_1^{(0)}}{H_{10}}\right]
 =
  C\left({\bf b}_1^{(0)}\cdot\nabla +
\partial_x b_{x1}^{(0)} + \partial_y b_{y1}^{(0)}
\right)
\left(\partial_x b_{y1}^{(0)} - \partial_y b_{x1}^{(0)}\right),
\eeqa
while for the top layer we have
\beqa
& & \left(\partial_t + {\bf u}_2^{(0)}\cdot\nabla \right)
\left[\partial_x v_2^{(0)} - \partial_y u_2^{(0)} - 2\frac{h_2^{(0)}-h_1^{(0)}}{H_{20}-H_{10}}\right]
 =
 \nonumber \\ & & \hskip 3.25cm
  C\left(\frac{\rho_1}{\rho_2}\right)\left({\bf b}_2^{(0)}\cdot\nabla +
\partial_x b_{x2}^{(0)} + \partial_y b_{y2}^{(0)}
\right)
\left(\partial_x b_{y2}^{(0)} - \partial_y b_{x2}^{(0)}\right).
\eeqa
In order to simplify the expressions appearing above we rewrite
the heights $h_i^{(0)}$ in terms of the streamfunctions
\beq
h_1^{(0)} = \frac{2}{{\rm Bu}^2}
\left(
\frac{\psi_1^{(0)} - \frac{\rho_2}{\rho_1}\psi_2^{(0)}}
{1 - \frac{\rho_2}{\rho_1}} \right), \qquad
h_2^{(0)} = \frac{2}{{\rm Bu}^2}\psi_2^{(0)}.
\eeq
We may now rewrite the equations in more transparent form
\beqa
& & \left(\partial_t + {\bf u}_1^{(0)}\cdot\nabla \right)Q_1^{(0)}
+\beta v_1^{(0)} =
  C\left({\bf b}_1^{(0)}\cdot\nabla +
q_{m1}^{(0)}
\right) J_1^{(0)}, \\
& & \left(\partial_t + {\bf u}_2^{(0)}\cdot\nabla \right)Q_2^{(0)}
+\beta v_2^{(0)} =
  C\left(\frac{\rho_1}{\rho_2}\right)\left({\bf b}_2^{(0)}\cdot\nabla +
q_{m2}^{(0)}
\right) J_2^{(0)}
\eeqa
where the potential vorticity in each layer is defined by
\beqa
Q_1^{(0)} &=& \nabla^2 \psi_1^{(0)} - \frac{1}{L_{10}^2}
\left(\psi_1^{(0)} - \frac{\rho_2}{\rho_1}\psi_2^{(0)}\right), \label{Q_1_not_def} \\
Q_2^{(0)} &=& \nabla^2 \psi_2^{(0)} - \frac{1}{L_{21}^2}
\left(\psi_2^{(0)} - \psi_1^{(0)}\right),
\eeqa
in which the $L_{10}$ and $L_{21}$ denote the Rossby radius
of deformation for layers 1 and 2 respectively,
\beq
\frac{1}{L_{10}^2} \equiv
\frac{4}{{\rm Bu}^2 H_{10}\left(1 - \frac{\rho_2}{\rho_1}\right)},
\qquad
\frac{1}{L_{21}^2} \equiv
\frac{4}{{\rm Bu}^2 (H_{20}-H_{10})\left(1 - \frac{\rho_2}{\rho_1}\right)}.
\eeq
The form presented above diverges slightly from the form presented in Vallis wherein
the upper layer is constrained by an upper boundary while here we allow for the upper
layer to undulate freely.  The difference then is that the ratio $\rho_2/\rho_1$
appearing on the RHS of (\ref{Q_1_not_def}), defining $Q_1^{(0)}$,  would be replaced by $1$.
We have also defined the currents $J_i^{(0)}$ appropriate to layer $i$
\beq
J_1^{(0)} \equiv \partial_x b_{y1}^{(0)} - \partial_y b_{x1}^{(0)},
\qquad
J_2^{(0)} \equiv \partial_x b_{y2}^{(0)} - \partial_y b_{x2}^{(0)},
\eeq
and the corresponding layer densities $q_{mi}^{(0)}$,
\beq
q_{m1}^{(0)} \equiv \partial_x b_{x1}^{(0)} + \partial_y b_{y1}^{(0)}, \qquad
q_{m2}^{(0)} \equiv \partial_x b_{x2}^{(0)} + \partial_y b_{y2}^{(0)}.
\eeq
All the terms in these equations, as appearing in the summary Section 6, will
be expressed with
their individual superscripts ``$(0)$" removed.
 Additionally, the expression for Bu
is rewritten in terms the definition of the Rossby Deformation Radii.

\section{Further development of the magnetostrophic limit and the derivation of a closed set of equations}\label{MSeqns}

The leading order magnetostrophic balance (\ref{lowest_order_magnetostrophic_balance}) leads
to a diagnostic specification of the horizontal velocities which we explitly write here
\beqa
2v_i^{(0)} &=& -\left(\frac{{\tilde{{\rm Bu}}}^2}{\rho_i}\partial_x \hat\Pi_i^{(0)} +
\frac{{\rm A}}{\rho_i}{\bf b}_i^{(0)}\cdot \nabla b_{xi}^{(0)}\right),
\label{MS_vi}
 \\
2u_i^{(0)} &=&  \frac{{\tilde{{\rm Bu}}}^2}{\rho_i}\partial_y\hat\Pi_i^{(0)} + \frac{{\rm A}}{\rho_i}{\bf b}_i^{(0)}\cdot \nabla b_{yi}^{(0)}.
\label{MS_ui}
\eeqa
We shall assume henceforth that the horizontal components of the velocity and magnetic
fields are z-independent within a given layer.  Thus,
the leading order induction equation is the same as (\ref{leading_order_induction}) which
we rewrite here for convenience,
\[
\left(\partial_t + {\bf u}_i^{(0)}\cdot \nabla\right) {\bf b}_i^{(0)}
=
{\bf b}_i^{(0)}\cdot\nabla {\bf u}_i^{(0)}.
\]
The horizontal velocity components are expressed in terms of ${\bf b}_i^{(0)}$ and the perturbation
height fluctuations
$h_i$ are contained in the perturbation pressure fields $\hat \Pi_i^{(0)}$.
Because the horizontal divergence of the lowest order velocity field was shown to be non-zero and subsequently
assumed to be independent of the vertical coordinate $z$, we can write the leading order vertical velocity
to be
\[
w_i^{(0)} =w_{i0}^{(0)}+ z \Omega_i^{(0)},
\]
where, as before, $w_{i0}^{(0)}$ is a z-independent undetermined vertical velocity.
Because of the relationships set up between the velocities according to the
leading order continuity equation
\[
\partial_x u_i^{(0)} + \partial_x v_i^{(0)} + \partial_x w_i^{(0)} = 0,
\]
together with the help of (\ref{horizontal_divergence_MS}),
we can express the quantity
$\Omega_i^{(0)}$ in terms of the horizontal magnetic field quantities as
\beq
\Omega_i^{(0)} =
-\frac{{\rm A}}{2\rho_i}\left(\partial_x b_{xi}^{(0)} +
\partial_x b_{yi}^{(0)} +
{\bf b}_i^{(0)}\cdot\nabla\right)J_i^{(0)}.
\eeq
We develop
an equation for the height fluctuations in the same manner as was done in the development of the MGQ equations.     Because we have argued
the fluctuations are on the same scale as the mean height of any given layer (i.e. because $\delta = \order 1$), instead of expressing
the height $H_i$ as the sum $H_{i0} + \delta h_i$ we shall simply stick with the expression $H_i$
since it now makes no difference as both quantities are the same order of magnitude.
Thus the vertical motion of layer height with position $z=H_i$ as viewed from the layer with index $i$
is to leading order
\[
\left(\partial_t + {\bf u}_i^{(0)}\cdot\nabla\right) H_i = w\bigr|_{z=H_i} = w_{i0}^{(0)}+ H_i \Omega_i^{(0)}.
\]
Similarly the motion of the layer height with position $z=H_{i-1}$ as viewed from the same layer with index $i$
is to leading order
\[
\left(\partial_t + {\bf u}_i^{(0)}\cdot\nabla\right) H_{i-1} = w\bigr|_{z=H_i} = w_{i0}^{(0)}+ H_{i-1} \Omega_i^{(0)}.
\]
Subtracting these two equations and reordering the results reveals
\beq
\left(\partial_t + {\bf u}_i^{(0)}\cdot\nabla\right)\ln\left(H_i - H_{i-1}\right) = \Omega_i^{(0)} =
-\frac{{\rm A}}{2\rho_i}\left(\partial_x b_{xi}^{(0)} +
\partial_x b_{yi}^{(0)} +
{\bf b}_i^{(0)}\cdot\nabla\right)J_i^{(0)}.
\label{Hi_MS}
\eeq
As presented, this magnetostrophic limit involves evolving equations
(\ref{horizontal_divergence_MS}) and (\ref{Hi_MS}) with the horizontal flow quantities
determined by the magnetstrophic balances explicitly
given in (\ref{MS_vi}) and (\ref{MS_ui}), where the perturbation pressure fields
$\Pi_i^{(0)}$, given in (\ref{hat_pi_definition}),
are re-expressed in terms of $H_i$.  We remind the reader that
(\ref{hat_pi_definition}) explicitly shows the pressure field quantities
for a two-layer system.  In general, for layers $i<N$ with $h_i$ expressed in terms
of $H_i$, we would have,
\beq
\hat\Pi_i^{(0)} = \rho_{i}H_{i} +
\sum_{k=i+1}^N \rho_k(H_k-H_{k-1}) ,
\eeq
and for layer $N$ we have $\hat\Pi_N^{(0)} = \rho_{N}H_{N}$.
\par
To illustrate how these equations appear we write them out for a single layer atmosphere where
we drop the superscripts,
\beqa
& &
\left(\partial_t + {\bf u}_1\cdot \nabla\right) {\bf b}_1 =
{\bf b}_1\cdot\nabla {\bf u}_1, \\
& & \left(\partial_t + {\bf u}_1\cdot\nabla\right)\ln H_1 =
-({{\rm A}}/{2})\left(\partial_x b_{x1} +
\partial_x b_{y1} +
{\bf b}_1\cdot\nabla\right)J_1,
\eeqa
in which the velocities and currents are diagnostically given by
\beqa
2v_1 &=& -{{\tilde{{\rm Bu}}}^2}\partial_x H_1 -
{\rm A}{\bf b}_1\cdot \nabla b_{x},
 \\
2u_1 &=& {{\tilde{{\rm Bu}}}^2}\partial_y H_1 +
{\rm A}{\bf b}_1\cdot \nabla b_{y},
\\
J_1 &=& \partial_x b_y - \partial_y b_x.
\eeqa
Note that these equations are governed by the two parameters, the modified Burger
number $\tilde{\rm Bu}$ and the Acheson number $A$.

\section{Summary and brief discussion of the equations of magnetoquasigeostrophy}
The result of the lengthly procedure detailed in Section \ref{detailed_MGQ}
 are summarized for two layers.  The equations for the potential vorticity in each layer
are given explicitly,
\beqa
& & \left(\partial_t + {\bf u}_1\cdot\nabla \right)Q_1
+\beta v_1 =
  \varphi_1C\left(q_{1}^{(m)}+{\bf b}_1\cdot\nabla
\right) J_1, \label{PVeq_1} \\
& & \left(\partial_t + {\bf u}_2\cdot\nabla \right)Q_2
+\beta v_2 =
  \varphi_2C\left(\frac{\rho_1}{\rho_2}\right)\left(q_{2}^{(m)}+{\bf b}_2\cdot\nabla
\right) J_2, \label{PVeq_2}
\eeqa
in which the horizontal velocities in vector form
are given by ${\bf u}_i = u_i \hat{\bf x} + v_i \hat{\bf y}$.
The gradient operator $\nabla$ is now understood to be two-dimensional
so that, for instance, ${\bf u}_i\cdot\nabla = u_i\partial_x + v_i\partial_y$.
The potential vorticities for each layer, $Q_i$, are given by
\beqa
Q_1 &=& \nabla^2 \psi_1 - \frac{1}{L_{10}^2}
\left(\psi_1 - \frac{\rho_2}{\rho_1}\psi_2\right),  \\
Q_2 &=& \nabla^2 \psi_2 - \frac{1}{L_{21}^2}
\left(\psi_2 - \psi_1\right),
\eeqa
where the streamfunctions $\psi_i$ relate to the velocity in each layer
according to
\beq
u_i = -\partial_y \psi_i, \qquad v_i = \partial_x \psi_i,
\eeq
and where the 2 dimensional Laplacian operator is given by $
\nabla^2 \rightarrow \partial_x^2 + \partial_y^2$.   Note that because
the leading order horizontal velocities are geostrophic it follows
that their horizontal divergences are zero, i.e.
\[ \nabla\cdot {\bf u}_i = 0.
\]
The (non-dimensionalized) Rossby radii of deformation for each layer are given by
\beq
{L_{10}^2} \equiv
\frac{g{\cal H}H_{10}}{4\Omega_0^2 {\cal L}^2} \left(1 - \frac{\rho_2}{\rho_1}\right),
\qquad
{L_{21}^2} \equiv
\frac{g{\cal H}(H_{20}-H_{10})}{4\Omega_0^2 {\cal L}^2} \left(1 - \frac{\rho_2}{\rho_1}\right),
\eeq
where $\rho_1$ and $\rho_2$ are the densities of each corresponding layer.
The dimensional value of each level height ${\cal H}_{i}$ are given respectively
according to their non-dimensional fractional measures $H_{i0}$. In other
words if ${\cal H}_{i0}$ represents the vertical coordinate where the transition
from density $\rho_i$ to $\rho_{i+1}$ occurs in steady state then the following
may be defined
$
H_{10} \equiv {{\cal H}_{10}}/{{\cal H}}, \ \ H_{20} \equiv {{\cal H}_{20}}/{{\cal H}}.
$
Since  only two layers are considered here the value of $H_{20} = 1$
as ${\cal H}_{20} = {\cal H}$ by construction.  It follows
that $H_{10}$ is some number less than 1.  The horizontal magnetic field
in each layer is denoted in vector form with
${\bf b}_i = b_{xi}\hat {\bf x} + b_{yi}\hat {\bf y}$
and the equations for their evolution are given by
\beqa
(\partial_t + {\bf u}_1\cdot \nabla){\bf b}_1 &=& ({\bf b}_1\cdot\nabla){\bf u}_1, \label{b1_eqn} \\
(\partial_t + {\bf u}_2\cdot\nabla){\bf b}_2 &=& ({\bf b}_2\cdot\nabla){\bf u}_2.
\label{b2_eqn}
\eeqa
As in the study considered by Gilman (2000), and unlike what is encountered for
the horizontal velocity components,
the horizontal magnetic field components
are not automatically divergence-free.  Of course, the source free condition
is met in three dimensions so that a non-zero divergence of the horizontal magnetic
field components will result in the generation of a linearly varying (with respect to
 the vertical coordinate) vertical magnetic field $b_{zi}$, as expressed in
  relationship (\ref{bz0_relation}).
It so happens that for these orderings the vertical field (if there is one
dynamically, see below) effects the
dynamics diagnostically through the pseudo-source term $q_{i}^{(m)}$ defined by
\beq
q_{i}^{(m)} \equiv \nabla\cdot{\bf b}_i = \partial_x b_{xi} + \partial_y b_{yi},
\eeq
and its influence can be seen in the Lorentz terms in the equations for the
potential vorticity (\ref{PVeq_1}-\ref{PVeq_2}).
The source term $q^{(m)}$ is the  pseudo-magnetic monopole distribution
referred to in the Introduction.
  Finally the $z$-directed
currents for each
layer are $J_i$ and defined by
\beq
J_i \equiv \partial_x b_{yi} - \partial_y b_{xi}.
\eeq
\par
An equation for the source terms $q_i^{(m)}$ may be developed by operating
each equation (\ref{b1_eqn}) and (\ref{b2_eqn}) by the divergence operator
and making use of the fact $\nabla\cdot{\bf u}_i = 0$.  This procedure results
in,
\beq
\left(\partial_t + {\bf u}_i\cdot\nabla\right)q_i^{(m)} = 0.
\eeq
This equation says something important:
that if $q_i^{(m)}(x,y) = 0$ initially then it remains
zero for subsequent times.  When the case, the consequence of this
is that the horizontal magnetic field components
obey a divergence-free condition and the evolution of the magnetic field is strictly
two-dimensional derivable from a flux function.  Note also that $q_i^{(m)}(x,y) = 0$ solutions of the
MGQ equation set form a self-contained subspace of all solutions of the MGQ equations.
When examination is focused wholly on this subclass, i.e. when
\beq
q_1^{(m)} = q_2^{(m)} = 0,
\eeq
it automatically follows
that $\nabla\cdot{\bf b}_i = 0$.  Consequently,
the flux functions relating to the magnetic field components are.,
\beq
b_{x1} = -\partial_y \phi_1, \qquad b_{y1} = \partial_x\phi_{1},
\qquad b_{x2} = -\partial_y \phi_2, \qquad b_{y2} = \partial_x\phi_{2}.
\eeq
Instead of
solving the vector equations (\ref{b1_eqn}) and (\ref{b2_eqn}),
the following simpler scalar equations must solved,
\beqa
(\partial_t + {\bf u}_1\cdot \nabla) \phi_1 &=& 0, \label{phi1_eqn} \\
(\partial_t + {\bf u}_2\cdot \nabla) \phi_2 &=& 0. \label{phi2_eqn}
\eeqa
In terms of flux functions, then, the currents in each layer are
given by
\beq
J_1 = \nabla^2 \phi_1, \qquad J_2 = \nabla^2 \phi_2.
\eeq
Thus the evolution equations given by (\ref{PVeq_1},\ref{PVeq_2}),
 with $q_i^{(m)}$ set to zero, and (\ref{phi1_eqn},\ref{phi2_eqn}), together with all of the supporting ancillary definitions form the basis of the equations appropriate for this subclass.
These are the same equations analyzed by Gilman in 1967-8.
\par
\par
\medskip
In an upcoming series of publications, we study the linear and nonlinear response of these equations 
which are amenable to relatively unencumbered theoretical analyses using well-known techniques used in both the meterological and fluid dynamics literature for studies of the QG equations.  To this end,
we shall use as a guide and expand upon the work laid out by Gilman and collaborators.  Our initial investigations will focus upon analyzing the physical stability of the compact magnetic vortices observed to emerge in the simulations of Cho (2008) and to hopefully, as a result, understand the nature of those structures and answer why they may be manifesting in those simulations with the observed robustness reported.

\acknowledgements {This work was begun and completed when the Astronomy Unit of Queen Mary University of London was part of the School of Mathematical Sciences.  Thanks are given to J. Y-K. Cho for introducing the author to the problems surrounding magnetization of exoplanet atmospheres.}

\end{document}